\documentclass[12pt]{article}
\usepackage{mathpazo,charter}
\usepackage[scaled=0.875]{helvet}
\usepackage[bookmarks,bookmarksnumbered]{hyperref}
\usepackage{amsmath,XXXE}
\usepackage{graphicx,color}
\usepackage{booktabs,multirow}
\def\oa{\textsf{\slshape a}\mkern1mu}
\def\ad{\textsf{\slshape a}\mkern1mu{}^\dagger}
\def\oH{\textsf{\slshape H}\mkern1mu}
\def\sH{\mathscr{H}}
\def\oJ{\textsf{\slshape J}\mkern1mu}
\def\oK{\textsf{\slshape K}\mkern1mu}
\def\Ks{\skew6\vec{\textsf{\slshape K}\mkern1mu}^{\,2}}
\def\oP{\textsf{\slshape P}\mkern1mu}
\def\oQ{\textsf{\slshape Q}\mkern1mu}
\def\Qd{\textsf{\slshape Q}\mkern1mu{}^\dagger}
\def\so{\mathop{\frak{so}}\nolimits}
\def\osp{\mathop{\frak{osp}}\nolimits}

\let\1=\Ione
 \SfTitles %section titles more modest
 \allowdisplaybreaks
\begin{document}

\begin{center}

{\LARGE\sf\bfseries\boldmath
        Spectrum-Generating Superalgebra\\[1mm]
        for Linear Harmonic Oscillators
 }\\*[5mm]
  %\vfill
{\large\sf\bfseries 
              T.~H\"{u}bsch
}\\*[1mm]
{\small\it
      Department of Physics \&\ Astronomy, Howard University, Washington, DC\\[-1mm]
      Department of Physics, University of Central Florida, Orlando FL
  \\[-4pt] {\tt\slshape  thubsch\,@\,howard.edu}
 }\\[5mm]
  %\vfill
{\sf\bfseries ABSTRACT}\\[3mm]
\parbox{128mm}{\addtolength{\baselineskip}{-2pt}\parindent=2pc\noindent
We show that the Hilbert space of the standard linear harmonic oscillator is a complete orbit of the $\osp(2{,}1;2)$ spectrum-generating superalgebra, and that this is the smallest such algebraic structure. The ubiquitous appearance of the linear harmonic oscillator in virtually all domains of theoretical physics guarantees a corresponding ubiquity of appropriate generalizations of this spectrum-generating superalgebra.
}
\end{center}
 \vspace{2mm}

\section{Introduction and Main Result}
 \label{s:IRS}
The linear harmonic oscillator (LHO) is possibly the best known and most thoroughly described quantum-mechanical system; see any text such as Ref.\cite{rKSL-QTh}, to name a recent one.
 It is also well embedded in the standard framework of all branches of theoretical physics, and in particular field theory\cite{rAZ-QFT} and string theory\cite{rRS-StrTh}, to name but a few. Harmonic oscillators are even used as a case study in mathematical literature\cite{rWyb,rNH-GQ}.
 Spectrum-generating transformations relate states of different energies, but still generate algebraic structures (algebras and groups) very much akin to symmetries. The spectrum-generating algebra for the LHO and even its $n$-dimensional generalizations has been known for almost four decades\cite{rUN-SymLHO}; see also {\em\/Case Study I\/} in Ref.\cite{rWyb}.
 
 It will be shown herein, however, that this standard LHO---without any fermionic or matrix extension---actually admits a spectrum-generating {\em\/superalgebra\/}, which has the distinguished capability of re-generating the whole Hilbert space from any one particular state. On this point, the present analysis differs from the approach and results found elsewhere in the literature; see \eg\ Ref.\cite{rMP-SHO96}, or Ref.\cite{rHPV-SHO10} for a more recent account and further references. The spectrum-generating {\em\/superalgebra\/} of the standard, bosonic LHO presented herein does not include supersymmetries, \ie, fermion-boson exchanging transformations that commute with the Hamiltonian and the anticommutator of which contains the Hamiltonian. Consequently, the spectrum-generating superalgebra is manifestly compatible with the algebraic structures found in higher-dimensional spacetimes and corresponding results such as the spin-statistics theorem.

%:Puzzle
\section{The Spectrum-Generating Algebra}
 \label{s:Puzzle}
The Hamiltonian $\oH$ for a linear harmonic oscillator is given in the well-known form\ft{Virtually every text on quantum-mechanics will suffice for the most part, but see Ref.\cite{rKSL-QTh} for a recent one.}:
\begin{equation}
 \oH = \hbar\w(\ad\oa+\inv2),\quad
 \oH\ket{n}=E_n\ket{n},\quad
 E_n=\hbar\w(n{+}\inv2),\quad
 n=0,1,2,\ldots
 \label{e:LHO}
\end{equation}
where $\oa$ and $\ad$ are the annihilation and creation operators, that satisfy:
\begin{equation}
 \oa\ket{n}=\sqrt{n}\ket{n{-}1},\quad
 \ad\ket{n}=\sqrt{n{+}1}\ket{n{+}1},\quad
 [\oa,\ad]=\Ione.
 \label{e:LHOkets}
\end{equation}
Thus, the LHO Hilbert space
\begin{equation}
 \sH\Defl\{\ket{n}:\vev{k|n}=\d_{k,n},~\sum_n\ket{n}\bra{n}
  =\Ione,~n,k=0,1,2,\ldots\}
 \label{e:HLHO}
\end{equation}
is non-degenerate, and every LHO stationary state is unambiguously determined by the energy.

It is less well known\cite{rUN-SymLHO,rWyb} that the triple of operators
\begin{equation}
  \oK_-\Defl\inv2\oa\oa,\quad
  \oK_+\Defl\inv2\ad\ad,\quad
  \oK_3\Defl\inv2(\ad\oa+\inv2\Ione)=\inv{2\hbar\w}\oH,
 \label{e:LHO3}
\end{equation}
satisfy the relations:
\begin{equation}
 [\oK_3,\oK_\pm]=\pm\oK_\pm\quad\text{and}\quad
 [\oK_+,\oK_-]=-2\oK_3
 \label{e:so(2,1)}
\end{equation}
of the $\so(2{,}1)$ algebra.
 These relations are very similar to the familiar angular momentum algebra
except that $\oK_3$ (the rescaled Hamiltonian $\oH$) occurs with a negative sign in the right-hand side of the last equation\eq{e:so(2,1)}.

It is well known that in all quantum systems that admit a well-defined action of the angular momentum operators, the individual states (denoted $\ket{j,m}$) are uniquely specified only by providing both $m$ and $j$, where $m$ is the eigenvalue of $\oJ_3$, and $j$ determines the eigenvalues of $\skew6\vec{\oJ}^{~2}$ to be $j(j{+}1)$. These two quantifiers are related by the condition
 $|m|\leq j$ and $\triangle m\in\ZZ$,
which jointly restrict $m$ and $j$ to both be either integral or half-integral.

The 1-dimensional system\eqs{e:LHO}{e:HLHO} then admits the operator $\Ks=\fc12(\oK_+\oK_-{+}\oK_-\oK_+){-}\oK_3{}^2$, which commutes with $\oK_\pm$ and $\oK_3$ and also with $\oH$, so that the well-known LHO eigenstates, $\ket{n}$, must also be eigenstates of this $\Ks$. Yet, for each $n$ and $E_n$, there is only one state, $\ket{n}$ and the {\em\/single\/} quantum number, $n$, suffices to uniquely identify each state of the LHO. How come no additional quantum number ($\k$, the eigenvalue of $\Ks$) is needed to uniquely specify the LHO states?

The answer emerges from the direct computation:
\begin{eqnarray}
 \Ks
 &\Defl&\inv2\big(\oK_+\oK_-+\oK_-\oK_+\big)-\oK_3{}^2
  = \inv8(\oa\oa\ad\ad+\ad\ad\oa\oa)-\inv4(\ad\oa+\inv2\1)^2~,\nn\\
 &=&\inv8[2\ad\ad\oa\oa+4\ad\oa+2\1]-\inv4[\ad\ad\oa\oa+2\ad\oa+\inv4\1]
 ~=~\frc3{16}\1~,
 \label{e:K2=1}
\end{eqnarray}
proving that $\Ks\ket{n}=\frc3{16}\ket{n}$, $\forall n$, and where the non-negativity condition\cite{rKSL-QTh,rWyb}
\begin{equation}
 \big\|\,\oK_\pm\ket{n}\big\|^2
 =\vev{n|\skew6\vec{\oK}^{\,2}|n}+\vev{n|\oK_3(\oK_3{\pm}1)|n}
 =\frc3{16}+n(n{\pm}1)\geq0
 \label{e:k}
\end{equation}
is straightforwardly satisfied for all $n\in\ZZ$; note that $n=\pm\frc12$ are excluded by this inequality. Furthermore, states with $n<0$ turn out to have a negative norm (see, \eg,\cite{rKSL-QTh}), and the entire LHO Hilbert space forms a single, infinite-dimensional, representation of $\frak{so}(2{,}1)$, each element of which (state in the LHO Hilbert space) has the same eigenvalue of $\Ks$, $\k=\frc3{16}$, and differs only by the eigenvalue of $\oK_3$, $n=0,1,2,3,\dots$

Notice furthermore that $\oK_+,\oK_-$ and $\oK_3$ commute with the reflection $\oP:(\oa,\ad)\to(-\oa,-\ad)$, and that $\oP\ket{n}=(-1)^n\ket{n}$. Thus, the LHO Hilbert space, as a representation of $\frak{so}(2{,}1)$, naturally decomposes into the $\oP$-even and $\oP$-odd states
\begin{equation}
 \sH=\sH_+\oplus\sH_-~,\qquad
 \sH_\pm=\inv2(\Ione\pm\oP)\sH~,
 \label{e:H+H-}
\end{equation}
having symmetric and antisymmetric wave-functions in the, say, coordinate representation. Furthermore, direct iteration of\eq{e:LHOkets} produces
\begin{equation}
  \oK_+\ket{n}=\sqrt{(n{+}1)(n{+}2)}\ket{n{+}2}
   \quad\text{and}\quad
  \oK_-\ket{n}=\sqrt{n(n{-}1)}\ket{n{-}2}.
 \label{e:K+K-}
\end{equation}
It is now clear that the ladder operators $\oK_\pm$ can re-create:
\begin{itemize}\itemsep=-3pt\vspace{-2mm}
 \item all even states and so all of $\sH_+$ from any one even state $\ket{2k}$,
 \item all odd states and so all of $\sH_-$ from any one odd state $\ket{2k{+}1}$.
\end{itemize}
Thus, given a pair of states, $\ket{2k},\ket{2k'{+}1}$, for any $k,k'=0,1,2,\dots$, the $\frak{so}(2{,}1)$ algebra $\{\oK_\pm,\oK_3\}$ can generate the whole spectrum of states of the linear harmonic oscillator and is called the \textit{spectrum-generating algebra} of the LHO system\cite{rWyb}. In general, the utility of spectrum-generating algebras stems from their ability to provide considerable information about the Hilbert space|and even matrix elements by way of the Wigner-Eckardt theorem|without ever explicitly solving differential equations, computing integrals or even determining the complete energy spectrum.
 Note, however, that the LHO Hilbert space $\sH$ decomposes\eq{e:H+H-}, and cannot be re-created by $\so(2{,}1)$ from a single state: $\sH$ consists of two separate complete $\so(2{,}1)$-orbits.

\section{The Spectrum-Generating Superalgebra}
 \label{s:A}
Firstly, note that
\begin{equation}
 \oK_3=\inv2(\ad\oa{+}\inv2\Ione)=\inv4(\ad\oa{+}\oa\ad)
 =\inv4\{\ad,\oa\},
 \label{e:LHOH}
\end{equation}
which motivates considering, besides the \textit{commutator} binary operations~(\ref{e:so(2,1)}), also the \textit{anticommutator} relations:
\begin{equation}
 \{\oa,\ad\}\Defl(\oa\ad{+}\ad\oa)=4\oK_3,\quad
 \{\ad,\ad\}=2\ad{}^2=4\oK_+,\quad
 \{\oa,\oa\}=2\oa{}^2=4\oK_-.
 \label{e:A1}
\end{equation}
This prompts the definitions
\begin{equation}
  \oQ\Defl\inv{\sqrt2}\oa \quad\text{and}\quad \Qd\Defl\inv{\sqrt2}\ad.
% \label{e:}
\end{equation}

Direct computation then easily completes the algebraic relations:
\begin{gather}
 [\oK_3,\Qd]=+\inv2\Qd,\quad
 [\oK_3,\oQ]=-\inv2\oQ, \label{e:A2}\\
 [\oK_+,\Qd]=0,\quad
 [\oK_+,\oQ]=-\Qd,\quad
 [\oK_-,\Qd]=+\oQ,\quad
 [\oK_-,\oQ]=0, \label{e:A3}\\
  \{\oQ,\Qd\}=2\oK_3,\quad
 \{\Qd,\Qd\}=2\oK_+,\quad
 \{\oQ,\oQ\}=2\oK_-.\label{e:A4}
\end{gather}
The operator $\oK_3$ generates the $\so(2)\subset\so(2{,}1)$ maximal compact subalgebra. In a real three-dimen\-sional space $(v_1,v_2,v_3)$ where $\so(2{,}1)$ linear transformations preserve the pseudo-Eucli\-dean quad\-ra\-tic form $v_1^{\,2}{+}v_2^{\,2}{-}v_3^{\,2}$, the operator $\oK_3$ generates rotations in the $(w_1,w_2)$-plane. Therefore, the results\eq{e:A2} imply that the pair $(\oQ,\Qd)$ spans the 2-dimensional spin-$\fRc12$ representation of the $\so(2{,}1)$ algebra, and $(\oQ,\Qd)$ are ``rotated'' into each other by $\oK_\pm$; see\eq{e:A3}.

 The collection of relations\eq{e:so(2,1)} and\eqs{e:A2}{e:A4} thus proves that the LHO operators
\begin{subequations}
 \label{e:so(2,1|2)}
\begin{gather}
  \oK_+    \Defl\inv4\{\ad,\ad\},\quad
  \oK_-    \Defl\inv4\{\oa,\oa\},\quad
  \oK_3    \Defl\inv4\{\ad,\oa\},\label{e:so(2,1}\\*
  \oQ      \Defl\inv{\sqrt2}\oa,\quad
  \oQ^\dag \Defl\inv{\sqrt2}\ad\label{e:|2)}
\end{gather}
\end{subequations}
span the structure of a \textit{superalgebra}, wherein the binary (graded) ``bracket'' operation is an anticommutator if both entries are odd powers of $\oa$ and $\ad$, and a commutator otherwise. Within this superalgebra, $\oK_+,\oK_-$ and $\oK_3$ span the even subalgebra, $\so(2{,}1)$, while the operators\eq{e:|2)} span the odd part. Amusingly thus, although the operators $\oa$ and $\ad$ are bosonic\ft{The $\oa$ and $\ad$ do not square to zero as fermionic, anticommuting annihilation and creation operators would.}, the superalgebra relations\eq{e:so(2,1)} and\eqs{e:A2}{e:A4} use anticommutators of odd powers of $\oa$ and $\ad$, precisely as if they were fermionic, anticommuting annihilation and creation operators.
 The superalgebra generated by the operators\eq{e:so(2,1|2)} extends $\so(2{,}1)$ by including the spin-$\fRc12$ {\em\/doublet\/} $(\oQ,\Qd)$, and is denoted herein $\osp(2{,}1;2)$ to denote this extension\cite{rBDW}. This superalgebra is also labeled as $\mathfrak{osp}(2|1)$ and $\mathfrak{osp}(1/2)$; see, \eg, Refs.\cite{rMP-SHO96,rHPV-SHO10} and \cite{rKNT-BCH}, respectively.

Finally, as well known, the operators\eq{e:|2)} act on the stationary states of the LHO as specified in relation\eq{e:LHOkets}, and can re-create the entire Hilbert space\eq{e:HLHO} from any one state. The
 LHO Hilbert space\eq{e:HLHO} is therefore a complete orbit of the $\osp(2{,}1;2)$ superalgebra\eq{e:so(2,1|2)},
 which is therefore the {\em\/spectrum-generating superalgebra\/} of the (1-dimensional) linear harmonic oscillator.
 
 Note that the the spectrum-generating algebra $\so(2{,}1)$ cannot re-create the entire Hilbert space from only one state, whereas the spectrum-generating superalgebra $\osp(2{,}1;2)$ can.
 By the same token, the two sectors\eq{e:H+H-} are not mixed by $\so(2{,}1)$, but are mixed by $\osp(2{,}1;2)$.

\section{Necessity}
Whereas the 5-generator superalgebra $\osp(2{,}1;2)$ can indeed re-create the entire LHO Hilbert space from any one state, is it the smallest superalgebra that can do so?

It is well known that the ladder operators $\oa$ and $\ad$ suffice to that end. Most every quantum mechanics textbook cite the {\em\/commutator\/} algebra
\begin{equation}
  \oa,\ad:\quad [\oa,\ad]=\Ione
 \label{e:H1}
\end{equation}
as the simplest case of the Heisenberg algebra. This however does not include the LHO Hamiltonian, which is of course essential in all physics applications. The minimal spectrum-generating algebraic structure should include both
\begin{enumerate}\itemsep=-3pt\vspace{-2mm}
 \item the generators that can re-create the entire LHO Hilbert space from any one state,
 \item and the Hamiltonian of the LHO, so as to uniquely identify each LHO state.
\end{enumerate}

At the very least, such an algebraic structure must include the generators
\begin{subequations}
 \label{e:MSG}
\begin{equation}
 \oK_3 = \inv{2\hbar\w}\oH,\quad
 \oQ=\inv{\sqrt2}\oa,~~\Qd=\inv{\sqrt2}\ad, \label{e:MSG1}
\end{equation}
which satisfy
\begin{equation}
 \{\oQ,\Qd\}=2\oK_3,\quad
 [\oK_3,\oQ]=-\fc12\oQ,~~
 [\oK_3,\Qd]=+\fc12\Qd. \label{e:MSG2}
\end{equation}
\end{subequations}
Owing to the necessary use of the anticommutator in the first of the relations\eq{e:MSG2}, this is then necessarily a superalgebra, where $\oQ$ and $\Qd$ are odd generators and $\oK_3$ is even. For closure, however, we then must include the results
\begin{equation}
  \{\oQ,\oQ\}=2\oK_-\neq0
   \quad\text{and}\quad
  \{\Qd,\Qd\}=2\oK_+\neq0,
\end{equation}
which force the inclusion of the $\oK_\pm$ generators, and then also the relations\eq{e:A3} and finally also the relations\eq{e:so(2,1)}.

The 5-generator superalgebra $\osp(2{,}1;2)$ generated by the operators\eq{e:so(2,1|2)} that satisfy the relations\eq{e:so(2,1)} and\eqs{e:A2}{e:A4} has thus been fully reconstructed and is therefore the minimal spectrum-generating superalgebra of the linear harmonic oscillator.

Alternatively, the purely commutative algebraic structure
\begin{equation}
 \oQ,~\Qd,~\oK_3,~\Ione:\quad
 [\oQ,\Qd]=\inv2\Ione,~~
 [\oK_3,\oQ]=-\inv2\oQ,~~
 [\oK_3,\Qd]=+\inv2\Qd,
\end{equation}
is indeed simpler (all other commutators vanish) than\eq{e:so(2,1)} together with\eqs{e:A2}{e:A4}, but has the structure of the Heisenberg algebra\eq{e:H1}, extended by the {\em\/ad hoc\/} inclusion of the operator $\oK_3$ acting upon the Heisenberg algebra as its module.

By contrast, the even generators\eq{e:so(2,1} of the subalgebra $\so(2{,}1)\subset\osp(2{,}1;2)$ are produced as quadratic expressions in terms of the odd generators, and in terms of the binary operation (the graded ``bracket'') of the superalgebra itself. The structure of $\osp(2{,}1;2)$, with the generators\eq{e:so(2,1|2)} satisfying\eq{e:so(2,1)} and\eqs{e:A2}{e:A4}, is thus forced by closure and so is the natural algebraic structure satisfying the two requirements itemized in the beginning of this section.

\section{Conclusions}
 \label{s:Conc}
Since the standard linear harmonic oscillator appears in virtually every branch of theoretical physics as one of the most widely used model, approximation and even foundational building block, the spectrum-generating superalgebra\eq{e:so(2,1|2)} and its appropriate generalizations is then expected to have the same ubiquity and similar utility.
In addition, the concepts of both spectrum-generating algebras and superalgebras have thus been shown to be fairly elementary and approachable, albeit not usually encountered in the courses wherein the LHO is a staple case study.

\bigskip\paragraph{Acknowledgments:}
 I should like to thank Costas Efthimiou for useful discussions, and the Physics Department of the Faculty of Natural Sciences of the University of Novi Sad, Serbia, for the recurring hospitality and resources.
 This work was supported by the Department of Energy through the grant DE-FG02-94ER-40854.

%\bibliographystyle{elsart-numX}
%\bibliography{Refs}
%\end{document}

\raggedright\end{document}